# Dynamics of Charge Transfer in Ordered and Chaotic Nucleotide Sequences

## N.S. Fialko[*], V.D. Lakhno[**]

*Institute of Mathematical Problems of Biology, Russian Academy of Sciences, Pushchino, Moscow Region, 142290, Russia*

**Abstract.** Charge transfer is considered in systems composed of a donor, an acceptor and bridge sites of (AT) nucleotide pairs. For a bridge consisting of 180 (AT) pairs, three cases are dealt with: a uniform case, when all the nucleotides in each strand are identical; an ordered case, when nucleotides in each DNA strand are arranged in an orderly fashion; a chaotic case, when (AT) and (TA) pairs are arranged randomly. It is shown that in all the cases a charge transfer from a donor to an acceptor can take place. All other factors being equal, the transfer is the most efficient in the uniform case, the ordered and chaotic cases are less and the least efficient, accordingly. The results obtained are in agreement with experimental data on long-range charge transfer in DNA.

The problem of charge transfer in biomacromolecules, in particular DNA, is now included in the most important issues of molecular biophysics and biochemistry, which is dealt with in a number of theoretical and experimental works [1]-[9].

The idea that a DNA may have conducting properties was put forward long ago. Soon after the double helical structure of DNA was determined, the suggestion was made that its regularly ordered array of nucleotides might provide a path for the conduction of electrical charge [10]. Each of two DNA strands may be viewed as a chain, each site of which is one of the four bases: adenine (A), guanine (G), cytosine (C) and thymine (T), having different ionization potentials and overlap integrals of neighboring bases.

Now it is generally recognized that a DNA molecule in an equilibrium state does not have any free charge carriers. Excess electrons (anion-radicals) or holes (cation-radicals) can be introduced into DNA either via photoexcitation upon exposure of the molecule to ultraviolet radiation or by chemical reactions, for example, by intercalating some special molecular complexes [1]-[3], [6]-[8]. A possibility of proton transfer in DNA is also considered [9]. However, of special interest is the study of electron and hole transfer along a chain of base pairs, since the motion of radicals through the DNA molecule resulting in its destruction may possibly play a crucial role in the processes of mutagenesis and carcinogenesis. The understanding of charge transport phenomena in DNA is central for the development of DNA-based molecular devices and the use of DNA as a component for molecular electronics circuitry.

Experimental studies of charge transfer over DNA fragments have become practicable only recently due to, on the one hand, progress in nano- and femto-second technology, and on the other hand, the development of biochemical methods of covalent binding of molecular complexes (which play the role of a donor and an acceptor) to a DNA fragment with the already known sequence. In most experiments electrons or holes have been created in a special way in DNA fragments with a known sequence of base pairs. The electron transport rate in this case has been calculated either from measurements of fluorescence quenching or from the analysis of the relative extent of damage produced in various sites of the DNA helix during the process of charge transfer. During the past few years tremendous experimental work has been done to elucidate the mechanism of charge transfer in DNA, however, the problem is still to be solved.

---

[*] fialka@impb.psn.ru
[**] lak@impb.psn.ru

Double helical DNA containing a stacked array of base pairs can be considered as a molecular analogue of a one-dimensional solid-state crystal. Like a solid-state, a DNA base pair stack provides a medium to facilitate charge transfer. However, due to base pair dynamical motions and sequence-dependent inhomogeneities in base pair properties, charge transfer in DNA differs considerably from that in solid-state crystal.

Below we consider various types of nucleotide sequences acting as intermediaries in charge transfer.

The charge transfer along a chain, composed of nucleotide pairs (sites), is described by the Hamiltonian:

$$H = H_e + T_K + U_P + H_C \quad (1)$$

$$H_e = \sum_i \alpha_i a_i^+ a_i + \sum_{i,j} \nu_{i,j}(a_i^+ a_j + a_j^+ a_i), \qquad \alpha_i = \alpha_i^0 + \alpha_i' \tilde{u}_i,$$

$$H_C = \frac{1}{2}\sum_{i,j} \gamma_{ij} \tilde{u}_i \tilde{u}_j, \qquad T_K = \frac{1}{2}\sum_i M_i \left(\frac{d\tilde{u}_i}{d\tilde{t}}\right)^2, \qquad U_P = \frac{1}{2}\sum_i k_i \tilde{u}_i^2.$$

Here $a_i^+, a_i$ are the operators of creation and annihilation of the excitation at the $i$-th site; $H_e$ is the operator of the excitation energy; $\nu_{ij}$ are matrix elements of the transition; $\alpha_i^0$ is the electron energy at the $i$-th site; $T_K$ is the kinetic energy of the sites; $M_i$ is the mass of the $i$-th site; $\tilde{u}_i$ is the displacement of the $i$-th site from the equilibrium position; $U_P$ is the potential energy of the sites; $k_i$ are elastic constants; $H_C$ is the coupling term [11]. If the coupling coefficients $\gamma_{ij} = 0$, the model is reduced to a single frequency (i.e. Einstein oscillator) model. The case $\gamma_{ij} \neq 0$ corresponds to the introduction of dispersion into the oscillations. It is believed that excitation energy $\alpha_i$ depends linearly on the site displacement, $\alpha_i'$ has the meaning of the coupling constant between quantum and classical subsystems; $i = 1,...,N$, $N$ is the number of sites in the chain (as in [5]).

The equations of motion in nondimensional variables, resulting from Hamiltonian (1) in the neighborhood approximation, have the form:

$$i\frac{db_i}{dt} = \eta_i b_i + \eta_{i,i+1} b_{i,i+1} + \eta_{i,i-1} b_{i,i-1} + \kappa_i \omega_i^2 u_i b_i, \quad (2)$$

$$\frac{d^2 u_i}{dt^2} = -\omega_i' \frac{du_i}{dt} - \omega_i^2 u_i - \xi_{i-1,i} u_{i-1} - \xi_{i,i+1} u_{i+1} - |b_i|^2. \quad (3)$$

where (2) is the Schroedinger equation for the amplitudes of the probability for excitation to occur at an $i$-numbered site, satisfying the normalizing condition $\sum_n |b_n(t)|^2 = 1$, and (3) are the classical equations for site motions. We introduce the damping term in classical equations, $\omega_i'$ is the friction coefficient.

The designations introduced in (2), (3) correlate to the parameters of the Hamiltonian (1) as:

$$\eta_i = \tau \frac{\alpha_i^0}{\hbar}, \quad \eta_{ij} = \tau \frac{\nu_{ij}}{\hbar}, \quad \omega_i^2 = \tau^2 \frac{k_i}{M_i}, \quad \kappa_i \omega_i^2 = \tau^3 \frac{(\alpha_i')^2}{M_i \hbar},$$

$$u_i = \beta_i \tilde{u}_i, \quad \beta_i = \tau^2 \frac{\alpha_i'}{M_i}, \quad t = \tau \tilde{t}, \quad \xi_{ij} = \gamma_{ij} \frac{M_i}{\tau^2},$$

where $\tau = 10^{-14}$ sec is the arbitrary typical time constant, $\beta_i$ is the typical size of $i$-th site oscillations.

We have modeled the dynamics of excitation transfer along a chain of sites, in which some initial and terminal sites with $\eta_n \neq 0$ served as a donor and an acceptor of such excitation.

In numerical experiments we considered the following cases.

**1. Uniform case.** All the bridge sites are identical.

For all the sites, except for the donor and the acceptor, the parameter values similar to [5] are taken: $\tau = 10^{-14}$ sec, $\nu_{n, n\pm 1} \sim 0.13$ eV. In dimensionless variables this value corresponds to $\nu_{n\pm 1, n} = 2$, we choose $\eta_n = 0$; the other parameters are: $\omega_n = 0.002$, $\omega'_i = 0.00001$, $\kappa_n = 2.5$.

In modeling the donor we changed the electron energy $\eta_D$, $D = 1, \ldots, 10$, the other parameters are the same as for the chain. At the moment $t = 0$ the excitation is believed to be localized at the donor. To model the acceptor we magnified hundred-fold the frequencies $\omega_A$ at 10 terminal sites and took the friction coefficients $\omega'_A$ to be of the order of one. The calculations were carried out for different values of $\eta_D$ ($D = 1, \ldots, 10$) at the donor, $\omega_A$, $\omega'_A$ and $\eta_A$ ($A = 191, \ldots, 200$) at the acceptor, and $\xi$ at sites.

Though the time of irreversible excitation transfer can differ widely for different parameters, the qualitative pictures of charge transfer are quite similar.

At the moment $t_0 = 0$ the charge is localized at the donor (usually we choose it to reside at the first site: $|b_1(0)|^2 = 1$ ). Then, during the time $t_1 \sim 1 \cdot 10^{-14}$ sec a ``hump'' forms (Fig. 1, graph b) and starts traveling over the chain at the velocity of $\sim 1.3 \cdot 10^7$ cm/sec, smearing out.

Fig.1 b–e show the dynamics of the excitation motion over the polynucleotide fragment during the first 0.8 psec. The excitation, having developed a wave package appearance (graph b, c) travels the entire length of the chain for $\sim 0.5$ psec, then it reflects (graph d) and runs back to the beginning of the chain. The ``hump'' becomes progressively lower and becomes blurred superimposing on its tail (graph e).

After several reflections the excitation virtually evenly smears out over all the sites (graph f) (we calculated probabilities at each site averaged over a short period of time) and persists in this state for rather a long time. Then the probability for the excitation to occur at the acceptor starts increasing and gradually becomes close to one (graphs $g - i$).

We considered chains consisting of 150, 200 and 250 sites. Except for the ten initial sites (locus of formation), which simulate the donor, and ten terminal sites, which simulate the acceptor (trap sites), all the other sites are considered to be identical. The results of the numerical experiments with the 200-site model look qualitatively similar to the results for the 150- and 250- site chains.

For $\xi = 0$ the dynamics of charge transfer depending on the initial conditions and the donor and acceptor parameters is described in [12].

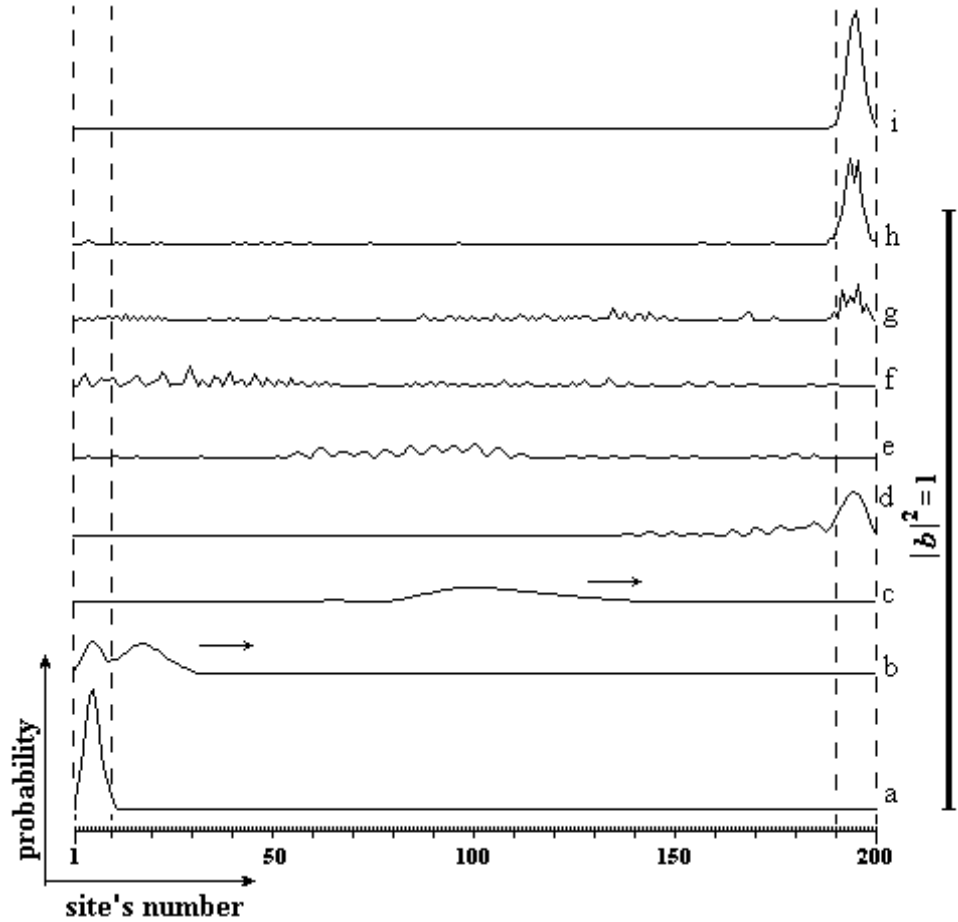

**Figure 1**. Graphs of the probabilities ($|b_n|^2_{t=const}$) on the sites at sequential time moments. The caliber segment of the probability is on the right. Dotted lines represent donor (on the left) and acceptor (on the right) sites. The arrows at graph b, c show the hump motion direction. The parameter values are: at donor $\eta_D = -0.4$, at acceptor $\eta_A = -0.4$, $\omega_A = 3$, $\omega'_A = 2$, $\xi = 0.000002 = \omega_{br}^2/2$, $\omega_{br}$ - nondimentional elastic constant of the bridge sites. At the initial moment the values of the probability amplitude at the donor are expressly preset to be approximately equal to the values at the acceptor when transition has actually occurred: $b_i(0) \cong b_{200-i}$ ($t = 50$ psec), $i = 1,\ldots,10$.

At small values of $\xi$ ($|\xi| \le \omega_{br}^2/2$, $\omega_{br}$ is a nondimentional elastic constant of the bridge sites) similar for all the sites, the transfer time (here by the transfer time we mean the interval [0,T], upon passing of which the probability of the occurrence of the excitation at the acceptor amounts to an average of no less than 0.9) changes only slightly as compared to the case $\xi = 0$.

**Table 1.** Time dependence of charge transfer on acceptor frequencies $\omega_A$ and friction coefficients $\omega'_A$ ($A = 191,\ldots,200$)

| $\omega_A$ | $\omega'_A$ | $T, 10^{-12}$sec $\xi = 0$ | $T, 10^{-12}$sec $\xi = 0.000001$ | $T, 10^{-12}$sec $\xi = 0.000002$ |
|---|---|---|---|---|
| 2 | 2 | 27.1 | 27.6 | 27.6 |
| 2 | 3 | 26.8 | 26.2 | 26.8 |
| 3 | 2 | 12.3 | 12.3 | 12.3 |
| 10 | 2 | 49.0 | 50.7 | 59.6 |

Table 1 lists the transfer time $T$ dependence on the values of the acceptor parameters and the value of coupling coefficient $\xi_n$, n=1,…,200 (on all the sites); the meaning of the other parameters corresponds to those in Fig. 1 (we chose the parameter values from clearness considerations).

Also we changed $\xi$ values on the donor and acceptor sites only.

On the basis of a large number of computer experiments, the following can be assumed.

As compared to the case $\xi = 0$:

- At small values of $\xi$ ($|\xi| \leq \omega_{br}^2/2$, $\omega_{br}$ is the nondimentional elastic constant of the bridge sites) the transfer time usually slightly increases (in general by 1–2%).

- At small values of $\xi$ at the donor and acceptor only ($\xi_{br} = 0$, $br = 11,…,190$) the transfer time decreases insignificantly (by less than 1%).

- At large positive values of $\xi_A$ at the acceptor ($\xi_A$, $\omega_A$, $\omega_A$ is a nondimentional elastic constant of the acceptor sites) the transfer time increases (up to 30%).

- At negative $\xi_A$ at the acceptor such that $-\xi_A \sim \omega_A$, $\omega_A$ the transfer time decreases (up to 10%)

- At negative $\xi_A$ such that $-\xi_A \sim \omega_A^2/2$, the transfer time increases several-fold (up to 700%).

Note that the patterns of transfer (Fig.1) are qualitatively similar for all physically meaningful $\xi$ values.

The time during which the trapped state is formed, i.e. the time of transfer, has weak dependence on the transfer distance and is almost entirely determined by the acceptor parameters and the initial state at the donor. The results of the numerical experiments with the 150-, 200- and 250-site chains confirm the scenario of the long-range electron transfer in DNA:

1) Oxidation of the acceptor has weak dependence on the distance [11]. For example, a weak dependence of the transfer rate on the sequence length was demonstrated in experiments by B.Giese et al. [13] for a nucleotide chain of (AT) pairs (uniform case).

2) There is a wide scatter in the transfer time as the experimental conditions change [14].

3) The transfer takes place only when the electron energies at the donor, the bridge and the acceptor have close values [15].

These results are in good agreement with the views on the energy control over the transfer rate which hold that the energy characteristics of the donor, the bridge and the acceptor should be correlated and close in values [15].

**2. Ordered case.** A chain of 180-site length composed of (TA) pairs with (AT) pairs embedded after each $N$-1 site is considered as a bridge. In other words we deal with chains of the type of $\left( \underbrace{T...T}_{N-1} A \right)_n$.

The difference between ionization potentials at thymine and adenine is $\Delta E \approx 0.21$ eV [16].

Accordingly, in dimensionless variables we took the energy values of bridge sites $\eta_k = 0$ for thymine and $\eta_k = -3$ for adenine. All the other parameters are the same as in the uniform case.

We dealt with the cases $N = 2$, $N = 5$ and $N=10$, i.e. with the chains of the types of $\begin{pmatrix} TA \\ AT \end{pmatrix}_n$, $\begin{pmatrix} TTTTA \\ AAAAT \end{pmatrix}_n$ and $\begin{pmatrix} TTTTTTTTA \\ AAAAAAAAT \end{pmatrix}_n$. We have not succeeded to find the parameter values at the donor and the acceptor at which after a lapse of rather a long time the probability at the acceptor would not be less than 0.9. Table 2 lists the distribution of the probabilities between the donor, the acceptor and the bridge sites (denoted as D, A, and Br, respectively) for there cases after a lapse of rather a long time (500 psec.). At the donor, $\eta_D = -1.5$, all the other parameters are the same as for the bridge sites; at the acceptor, $\eta_A = -1.5$, $\omega_A = 3$, $\omega'_A = 2$.

**Table 2.** Distribution of the probabilities between the donor, the bridge, and the acceptor after 500 psec

| $\xi_I$, $I = 1,\ldots,200$ | $N = 2$ | | | $N = 5$ | | | $N = 10$ | | |
|---|---|---|---|---|---|---|---|---|---|
| | D | Br | A | D | Br | A | D | Br | A |
| 0 | 0.47 | 0.47 | 0.06 | 0.27 | 0.31 | 0.42 | 0.12 | 0.15 | 0.73 |
| 0.000002 | 0.33 | 0.66 | 0.01 | 0.09 | 0.30 | 0.61 | 0.18 | 0.10 | 0.72 |

It is seen from the table that at $N = 2$ there is little probability that the transfer may occur; in the other cases partial transfer takes place, the larger being $N$ (i.e. the grid cell), the greater being the probability at the acceptor (for example, in the limit case $N = 90$, i.e. when one adenine resides in the middle of a thymine bridge and $\eta_D = \eta_A = -0.4$, on ~ 21.9 psec the probability for a charge to occur at the acceptor is not less than 0.9). So, in the ordered case the transfer depends not only on the donor and acceptor parameters, but also on the properties of the bridge itself.

Note that in the cases of $\begin{pmatrix} TTTTA \\ AAAAT \end{pmatrix}_n$ and $\begin{pmatrix} TTTTTTTTA \\ AAAAAAAAT \end{pmatrix}_n$ chains, a hole travels mainly over thymines, notwithstanding the fact that adenines have a lower ionization potential. The reason is that the matrix elements of a transition between thymines in one chain are several times greater than those for a transition between adenines [17].

**3. Chaotic case.** As opposed to the previous case, adenines are strewn chaotically over a chain of thymines.

We carried out calculations for two variants: the number of adenines made up on the average 20% and 10% of the total number of base pairs (which corresponds to $N = 5$ and $N = 10$ in the case 2). The relevant values of electron energies at the bridge sites were preset randomly with appropriate probability (e.g. for $N = 5$ the probability of $\eta_i = 0$ was 0.8, the probability of $\eta_i = -3$ was 0.2). We carried out 16 experiments for each variant. The mean values of the probability distributions calculated for 10 and 16 experiments are rather close.

For comparison purposes Table 3 lists the distribution of the probabilities between the donor, the bridge and the acceptor after 500 psec for $\xi = 0$ in uniform, ordered and chaotic chains (the parameter values are the same as in Table 2).

**Table 3.** Comparative list of the probability distribution after 500 psec

| | uniform | ordered $N = 10$ | chaotic 10% | ordered $N = 5$ | chaotic 20% |
|---|---|---|---|---|---|
| donor | 0.12 | 0.12 | 0.24 | 0.27 | 0.36 |
| bridge | 0.02 | 0.15 | 0.44 | 0.31 | 0.58 |
| acceptor | 0.86 | 0.73 | 0.32 | 0.42 | 0.06 |

This result demonstrates a possibility of long-range charge transfer in DNA with disordered sequence of nucleotide pairs which was observed experimentally [18].

The experiments on long-range (over 600 nm in λ–DNA) electron transfer [3], the transfer of a hole over the distance of 10 nm in a synthetic chain of (GC) nucleotide pairs [19] and long range hole transfer over 50 nucleotide pairs in [20] suggest that a charge transfer in DNA may occur at a large distance.

In conclusion we would like to suggest a possible explanation of the results of an experiment performed by Lewis et al. [21]. The overwhelming majority of experiments on charge transfer in DNA deal not with the dynamics of transfer but with its consequences, such as identification of oxidized DNA sites or molecular complexes bound to them. Accordingly, the transfer time is not determined. The only exception familiar to us is experiment [21] which deals with a hole transfer between (GC) sites separated by one (AT) pair. There the transfer time was found to be approximately equal to $10^{-6}$ sec. In our numerical experiments, the time of a transfer over a distance of 180 (AT) pairs was calculated to be much less than $10^{-9}$ sec. Therefore we speculate that the authors of [21] measured not the rate of a hole transfer between guanine sites (which makes up only fractions of a picosecond), but the rate of a reaction between a charge and water which is the lowest process in the transfer of a charge.

The work is supported by RFBR projects N 01–07–90317, 03–04–49255.